\documentclass[aps,prb,twocolumn,longbibliography]{revtex4-1}

\usepackage{epsfig}
\usepackage{soul}

\usepackage{bm}
\usepackage{graphicx}
\usepackage{amsmath}
\usepackage{amssymb}
\usepackage[utf8]{inputenc}
\usepackage[T1]{fontenc}
\usepackage{color}
\usepackage{upgreek}
\usepackage{subfigure}
\usepackage{placeins}
\usepackage{lipsum}
\usepackage{epsfig}
\usepackage[utf8]{inputenc}
\usepackage{wrapfig}

\usepackage[  colorlinks=true,  urlcolor=blue,  linkcolor=blue,  citecolor=blue]{hyperref}

\AtBeginDocument{%
    \newwrite\bibnotes
    \def\bibnotesext{Notes.bib}
    \immediate\openout\bibnotes=\jobname\bibnotesext
    \immediate\write\bibnotes{@CONTROL{REVTEX41Control}}
    \immediate\write\bibnotes{@CONTROL{%
    apsrev41Control,author="08",editor="1",pages="1",title="0",year="1"}}
     \if@filesw
     \immediate\write\@auxout{\string\citation{apsrev41Control}}%
    \fi
}%

\begin{document}

\title{Spin splitting in low-symmetry quantum wells beyond Rashba and Dresselhaus terms}

\author{G.\,V.\,Budkin} 
\author{S.\,A.\,Tarasenko}

\affiliation{Ioffe Institute, 194021 St.~Petersburg, Russia} 

\begin{abstract}
%We report on the study of the spin splitting of two-dimensional states in quantum wells grown along low-symmetry crystallographic axes based on cubic crystals.
%We show that beyond the Rashba and Dresselhaus contributions there is an additional term in the Hamiltonian of spin-orbit interaction, induced by structure inversion asymmetry, which couples the out-of-plane spin component with the in-plane momentum component.
%We carry out numerical calculations and develop a microscopic theory, which demonstrate that this interaction can dominate the $\bm k$-linear spin splitting of heavy-hole subbands.
%We report on the study of $\bm k$-linear spin splitting of two-dimensional states in quantum wells grown along low-symmetry crystallographic axes based on cubic crystals.
Spin-orbit interaction in semiconductor structures with broken space inversion symmetry leads to spin splitting of electron and hole states
even in the absence of magnetic field. We discover that, beyond the Rashba and Dresselhaus contributions, there is an additional type of the 
zero-field spin splitting which is caused by the interplay of the cubic shape of crystal unit cell and macroscopic structure asymmetry.
In quantum wells grown along low-symmetry crystallographic axes, this type of spin-orbit interaction couples the out-of-plane component of carrier's spin 
with the in-plane momentum while the coupling strength is controlled by structure inversion asymmetry. We carry out numerical calculations and 
develop an analytical theory, which demonstrate that this interaction can dominate $\bm k$-linear spin splitting of heavy-hole subbands.  
\end{abstract}

%\pacs{} 
\maketitle
  
\section{Introduction}
\label{introduction} 

Zero-field spin splitting of electron and hole subbands in semiconductors caused by spin-orbit (SO) interaction gives rise to a variety of exciting phenomena which are being explored for technological applications~\cite{Awschalom2007,Fabian2007,dyakonov_book}. Examples are the intrinsic and intrinsic inverse spin Hall effects~\cite{Sinova2015,Kimata2019}, the spin-galvanic effect~\cite{Ivchenko1990,Ganichev2002,Tarasenko2018,khokhriakov2020} and spin orientation by electric current~\cite{norman2014,Ganichev2020},  persistent spin helices~\cite{Schliemann2017,Kohda2017,Passmann2019}, and the paradigmatic spin field-effect transistor~\cite{Datta1990,Chuang2015}.
 
Two contributions to the zero-field spin splitting in zinc-blende-type quantum wells (QWs) are commonly considered: the Rashba and Dresselhaus terms~\cite{winkler_book,zawadzki2004,introduction_footnote}. 
The Rashba (or Bychkov-Rashba) SO interaction~\cite{Rashba1960,Vasko1979,Bychkov1984,winkler2000,manchon2015,Marcellina2017} is related to the QW structure inversion asymmetry (SIA) which can be tuned by an electric field applied along the QW normal.  
% The Rashba Hamiltonian has the form $H_R = \alpha (\bm \sigma \times \bm k)\cdot \bm n$,  
% where $\bm \sigma$ is the vector composed on the Pauli matrices and $\bm k$ is the in-plane wave vector, $\alpha$ is a constant for electrons 
% and $\propto k^2$ for heavy holes.
The effective magnetic field $\bm B_R$ corresponding to the Rashba splitting lies in the QW plane. 
% and,  in the isotropic approximation, $B_R \propto k$ and $\propto k^3$ for electrons and heavy holes, where  $\bm k $ is the electron/hole wave vector $\bm k$. 
The Dresselhaus term originates from bulk inversion asymmetry (BIA) of  
the underlying crystal~\cite{Dresselhaus1955,Dyakonov1986,Pikus1988,Rashba1988,Durnev2014}. The corresponding effective magnetic field $\bm B_D$ is tied to the crystallographic axes and, therefore, its direction depends on the QW crystallographic orientation~\cite{Dyakonov1986}. In particular, $\bm B_D$  lies in the QW plane in (001)-oriented structures whereas it points along the QW normal in (110)-oriented structures.
%{\color{blue} to footnote: Additional contributions to the SO splitting may come interface inversion asymmetry and strain. The former is typically small and plays a role if other sources of SO interaction are suppressed, like in symmetric Si/Ge QWs, or if electron/hole states are bound to interfaces, like in HgTe/CdHgTe topological insulators. Strain, particularly shear strain in lattice-mismatched (110)-grown QWs, can significantly increase SO splitting and make the Rashba splitting anisotropic.}
%{\color{red} no ref marcellina2017}

Technological achievements and the search for new physics stimulate the study of QWs grown along low-symmetry axes,
such as [012], [013], etc. While the analysis of atomic structure readily shows that the symmetry of such QWs is reduced
[down to the trivial $C_1$ group compared to the $C_{2v}$ group of (001)-grown QWs~\cite{ivchenko_book}] and, therefore, 
additional terms in the SO Hamiltonian get phenomenologically allowed, it is commonly assumed that the SO Hamiltonian still consists 
of the Rashba and Dresselhaus terms. The latter is obtained from the bulk Dresselhaus term by the proper projection onto 2D states 
in the coordinate frame relevant to the QW. 

Here, we show that there is an additional term in the Hamiltonian of zero-field splitting, which emerges in any QWs except those grown 
along the high-symmetry axes $\langle 001\rangle$, $\langle 011\rangle$, and $\langle 111\rangle$. The term is relevant for QWs 
based on cubic crystals, e.g., both zinc-blende-type and diamond-type crystals, and does not require bulk inversion asymmetry.
The effective magnetic field corresponding to this type of SO splitting points along the QW normal while its magnitude is controlled 
by structure inversion asymmetry and can be tuned by gate voltage. Moreover, we find that this SO interaction dominates 
the $\bm k$-linear spin splitting of heavy-hole (HH) subbands in low-symmetry QWs based on III-V semiconductors, 
such as (013)-grown GaAs QWs. In the framework of the Luttinger Hamiltonian,
we construct the effective Hamiltonian of the HH SO splitting for $(0lh)$-grown QWs, where $l$ and $h$ are the Miller indices, and study how 
the SO splitting depends on the QW orientation and width as well as the electric field applied along the growth direction.
Calculations also reveal that the $\bm k$-linear Rashba splitting of the HH states, which is absent in the isotropic approximation, 
emerges in the cubic model and has a strong dependence on the QW crystallographic orientation.

\section{Symmetry consideration}
\label{symmetry}
 
\begin{figure*}[!ht] 
	\centering
	\includegraphics[width=1.0\linewidth]{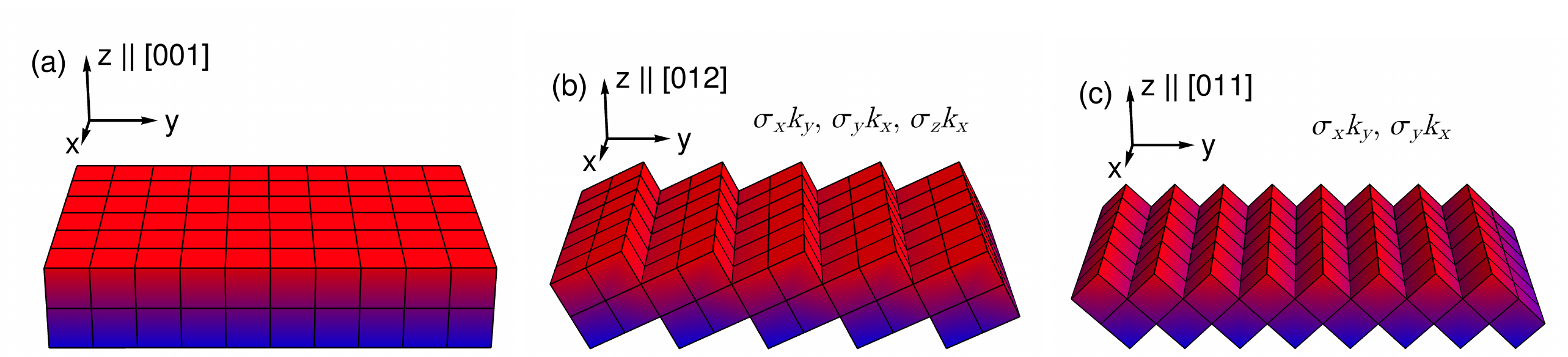}
	\caption{Sketch of the microscopic structures of QWs grown from a cubic crystal along the [001], [012], and [011] axes. 
	The cubic symmetry of crystal unit cells together with structure inversion asymmetry gives rise to $\bm k$-linear spin splitting of the HH subband in 
	(012)- and (011)-grown QWs with the terms shown in the figure.}
	\label{structure_sketch}     
\end{figure*}
 
First, we show that $\bm k$-linear SO splitting follows from symmetry consideration and construct the effective Hamiltonian.
Figure~\ref{structure_sketch} sketches the microscopic structures of asymmetric QWs grown from a cubic crystal along three different $[0lh]$ axes:
[001], [012], and [011].
The crystal unit cells are shown by cubes, the color gradient from blue to red illustrates the QW structure inversion asymmetry.
We focus on the cubicity of the unit cell and neglect its internal structure, in particular, bulk inversion asymmetry for zinc-blende-type crystals.
Moreover, for heavy holes in asymmetric quantum wells, the BIA term is typically smaller than the SIA term~\cite{Marcellina2017}.
% since it originates from a weak admixture of remote bands~\cite{Durnev2014,Marcellina2017}.   
   
Figure~\ref{structure_sketch} reveals that the symmetry of a QW depends on its crystallographic orientation if the cubicity of unit cells is taken into account. 
The highest symmetry, corresponding to the $C_{\text{4v}}$ point group in the cubic model, is achieved for (001)-grown QWs, Fig.~\ref{structure_sketch}a.
Such a symmetry allows for $\bm k$-linear Rashba splitting of conduction-band electrons (particles with the spin projections $\pm 1/2$ on the QW normal) and $\bm k$-cubic Rashba splitting of heavy holes (particles with the spin projections $\pm 3/2$). $\bm k$-linear Rashba splitting of HH is absent in this model and is known to be tiny in real (001)-grown zinc-blende-type QWs~\cite{winkler_book}.   
%~\cite{symmetry_footnote}.  
In (011)-grown QWs (Fig.~\ref{structure_sketch}c), the in-plane axes $x$ and $y$ get nonequivalent and the effective symmetry is reduced to the $C_{\text{2v}}$ point group. This symmetry enables $\bm k$-linear coupling of the $\pm 3/2$ states. 
As the results, the Rashba Hamiltonian for HH acquires $\bm k$-linear terms and becomes anisotropic.

QWs grown along any axes between [001] and [011], e.g., (012)-grown QWs as shown in Fig.~\ref{structure_sketch}b, has lower symmetry. In the cubic model, they are described by the $C_s$ point group with the only non-trivial symmetry element: the mirror plane $x \rightarrow -x$. In addition to the anisotropic Rashba term, the SO Hamiltonian in such QWs may contain the unusual term $\sigma_z k_x$ which is also invariant in the $C_s$ point group. Thus, the effective Hamiltonian of $\bm k$-linear SO interaction in $(0lh)$-grown QWs of the $C_s$ symmetry can be generally presented in the form
\begin{equation}
\label{so_symmetry}
 H_{\text{SO}}=\alpha_1 \sigma_y k_x-\alpha_2 \sigma_x k_y+\zeta \sigma_z k_x\:,
 \end{equation} 
 where $\alpha_1$, $\alpha_2$, and $\zeta$ are the SO coupling parameters, $\sigma_{x,y,z}$ are the Pauli matrices, $k_{x,y}$ are the in-plane wave vector components, and $x \parallel [100]$, $y \parallel [0h\overline{l}]$, and $z \parallel [0lh]$ is the coordinate frame relevant to $(0lh)$-grown QWs.
Below, we calculate these parameters for the HH subband and demonstrate that for most crystallographic orientations the term $\sigma_z k_x$ dominates. 
This term seems to be responsible for the out-of-plane spin polarization of surface states in (013)-grown HgTe/CdHgTe topological insulators observed in our numerical $\bm k$$\cdot$$\bm p$ calculations~\cite{dantscher2015}. %$
Note that similar term caused by BIA in  $(0lh)$-grown QWs would have the form $\sigma_z k_y$~\cite{Nestoklon2016}. 
  
%The results are summarized in Fig. 1

%"we find that for the single heterojunctions studied here the magnitudes of BIA terms are always
%much smaller than those of SIA terms." PHYSICAL REVIEW B 95, 075305 (2017)

\section{Microscopic theory}
\label{numerical}  
 
\begin{figure*}[!ht] 
	\centering  
	\includegraphics[width=1.0\linewidth]{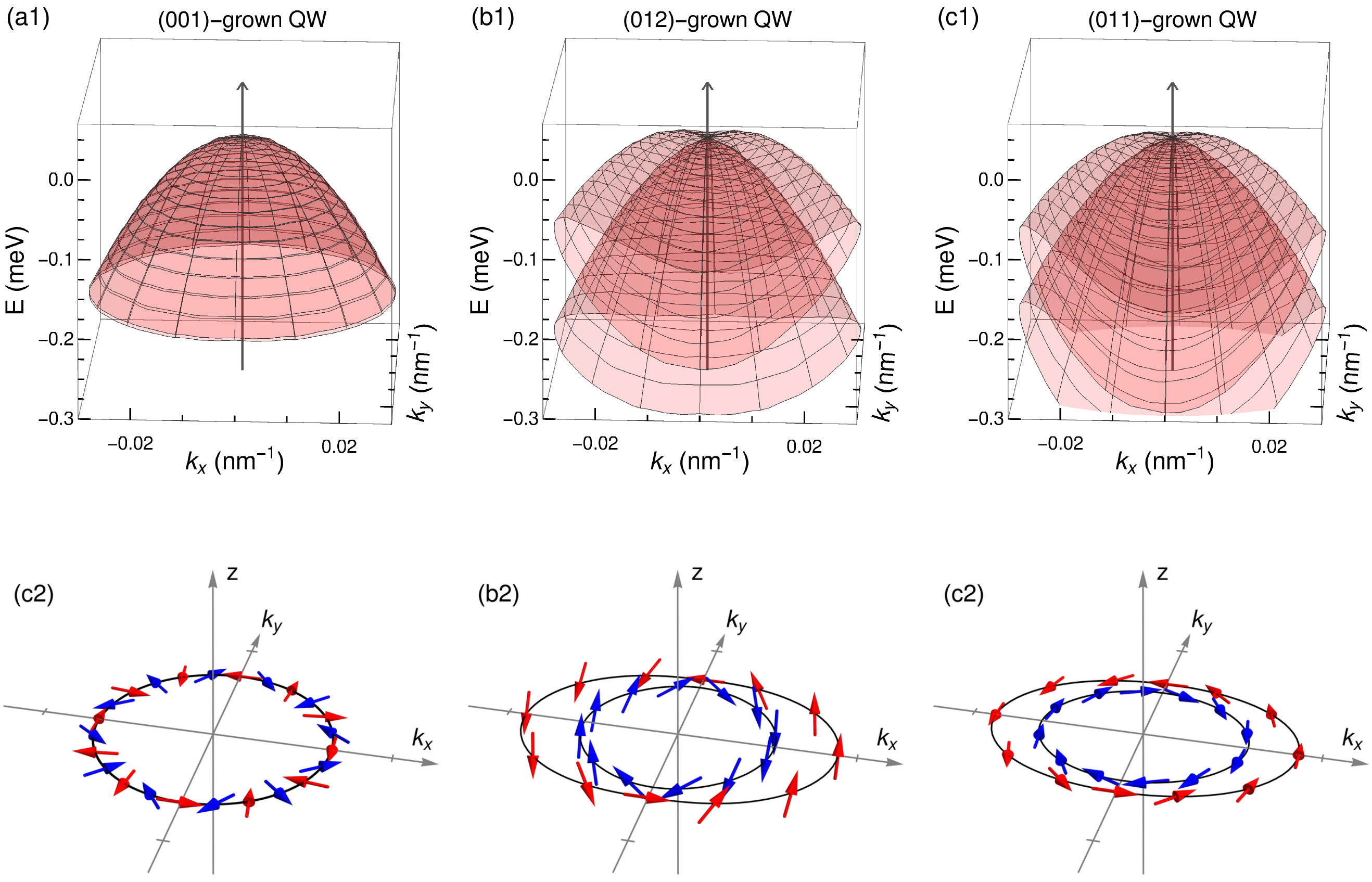}
	\caption{
(a1), (b1), and (c1) Energy spectra of ground heavy-hole subband in (001)-, (012)- and (011)-grown GaAs/AlAs QWs, respectively.
(a2), (b2), and (c2) Equipotential cross sections of the energy spectra. Red and blue arrows show the orientations of heavy-hole pseudospin 
in the spin subbands. Calculations are carried out for $5$-nm-wide QWs with the built-in electric field $E_z=400$~kV/cm.	
%
	%The results of $\bm{k}\cdot \bm{p}$ calculations for $5$~nm wide GaAs/AlAs quantum wells with different orientations.
	%(a) and (b) show the heavy-hole energy spectrum and pseudo-spin orientation for (001)-grown structure, respectively,
	%(c) and (d) are for (011)-grown QW and (e) and (f) are for (012)-grown QW.
	%Calculation are carried out for built-in electric field $E_z=400$~kV/cm.		 
	}
	\label{numerical_fig} 
\end{figure*} 

Now, we develop a microscopic theory of SO splitting. The $\Gamma_8$ valence band in bulk cubic semiconductors is described by the Luttinger Hamiltonian, which in the cubic axes has the form~\cite{ivchenko_book,winkler_book} 
\begin{multline}\label{H_L}
 H_L=
  \frac{\hbar^2}{2 m_0} \biggl[ -  \left( \gamma_1 + \frac52 \gamma_2\right) {\bm k}^2
  +  2 \gamma_2  ( \bm{J} {\bm k} )^2 \\
  + 2 (\gamma_3-\gamma_2)\sum_{i \neq j} (J_i k_i)(J_j k_j)\biggr]\:, 
 \end{multline} 
 where $\gamma_1$, $\gamma_2$, and $\gamma_3$ are the Luttinger parameters, $\bm k$ is the three-dimensional wave vector, and $\bm{J}$ is the vector composed of the matrices of the angular momentum 3/2. 
 % The last term in Eq.~\eqref{H_L} takes into account the cubic anisotropy of the energy spectrum.
  
To calculate the dispersion of valence subbands in $(0lh)$-grown QW, we rotate the Luttinger Hamiltonian in the QW coordinate frame $(x,y,z)$ and solve the Schr\"{o}dinger equation $H \Psi = E \Psi$ with the Hamiltonian
 \begin{equation}
 \label{full_H}
 H = H^{({\rm i})}_L + H^{({\rm a})}_L + eV(z) \:,
 \end{equation}  
 where $H^{({\rm i})}_L$ and $H^{({\rm a})}_L$ are the isotropic and anisotropic parts of the Luttinger Hamiltonian, respectively, and
 $eV(z)$ is the electrostatic potential energy~\cite{dantscher2015}. We use the simplest form of boundary conditions at the QW interfaces: 
 the continuity of $\Psi$ and $v_z \Psi$ at the interfaces, where $v_z = \hbar^{-1} \partial H_L / \partial k_z$. These boundary conditions applied to  
 $(0lh)$-grown structures correspond to the Hamiltonians 
 \begin{multline}
H^{({\rm i})}_L = \frac{\hbar^2}{2 m_0} \biggl[ - {\bm k}  \left( \gamma_1 + \frac52 \gamma_2 \right) {\bm k}
\, + 2 \sum_{ij} \{J_i J_j\}_s \, k_i \gamma_2 k_j \biggr] \:
\end{multline} 
%  
 %
% \begin{multline}
%H^{({\rm i})}_L = \frac{\hbar^2}{2 m_0} \biggl[ -{\bm k} \left( \gamma_1 + \frac52 \gamma_2\right) {\bm k} +\\ 
% 2 ( \bm{J}_\| \bm{k}_\|)  \gamma_2  ( \bm{J}_\| \bm{k}_\|)+
% 2 J_z  k_z \gamma_2 J_z k_z+\\ 4 \{J_x J_z\}_s \{\gamma_2,k_z\}_s k_x +4 \{J_y J_z\}_s \{\gamma_2,k_z\}_s k_y \biggr] \:
%\end{multline} 
%  
and   
\begin{multline}
H^{({\rm a})}_L = \frac{2\hbar^2}{m_0} \biggl\{ \{J_x J_y \}_s (\gamma_3 - \gamma_2) k_x k_y + \{J_x J_z \}_s \{ \gamma_3 - \gamma_2 , k_z \}_s k_x \\
+ \left[\{J_y J_z\}_s \cos 2\theta + \frac{J_z^2 - J_y^2}{2} \sin 2\theta \right] \\ 
\times \Biggl[\{\gamma_3-\gamma_2,k_z \}_s \, k_y \cos 2\theta \\
+ \frac{k_z (\gamma_3-\gamma_2) k_z  - (\gamma_3-\gamma_2)k_y^2}{2} \sin 2\theta \Biggr] \Biggr\} \:,    
\end{multline}
where $\bm k_\parallel = (k_x,k_y)$ is the in-plane wave vector, ${k_z = -i \partial_z}$, $\{A B\}_s=(AB+BA)/2$ is the symmetrized product, $\theta=\arctan(l/h)$ is the angle between the QW growth direction 
$[0lh]$ and $[001]$, and the Luttinger parameters may now depend on the $z$ coordinate.

\begin{table}
\begin{tabular}{ |c|c|c|c| } 
\hline
 & $\gamma_1$ & $\gamma_2$ & $\gamma_3$ \\
\hline 
GaAs & 6.98 & 2.06 & 2.93\\  
AlAs & 3.76 & 0.82 & 1.42 \\ 
\hline
\end{tabular}
\caption{Luttinger parameters of GaAs and AlAs~\cite{sodagar2009}.}
\label{gamma_table}
\end{table} 

The energy spectra of the ground HH subband in \mbox{(001)-,} (012)-, and (011)-grown GaAs/AlAs QWs are shown in Fig.~\ref{numerical_fig}.
The spectra are calculated using the Luttinger parameters listed in Table~\ref{gamma_table} and the valence band offset between GaAs and AlAs $590$~meV~\cite{wang2013}. The electrostatic potential $V(z)$ is set to constants in the barriers and $V(z) = E_z z$ inside the well.

Figures~\ref{numerical_fig}(a1), \ref{numerical_fig}(b1), and \ref{numerical_fig}(c1) reveal that the HH spin-orbit splitting dramatically depends 
on the QW crystallographic orientation. As expected, the splitting is negligible in (001)-grown QWs where $\bm k$-linear terms are absent and 
the Rashba splitting emerges only in the third order in the wave vector~\cite{winkler_book}. In contrast, anisotropic $\bm k$-linear splitting of the HH subband is clearly observed in the energy spectra of (012)- and (011)-grown QWs. Moreover, among all the QW orientations considered in Fig.~\ref{numerical_fig}, 
the splitting is maximal in the (012)-grown QW. This is most clearly seen in Figs.~\ref{numerical_fig}(a2), \ref{numerical_fig}(b2), and \ref{numerical_fig}(c2),
 where solid black lines show the equipotential cross sections of the energy spectra.  
 
To clarify the nature of the observed SO splitting we calculate the directions of pseudospin in the spin subbands. 
The pseudospin in the state $\Psi$ is defined as $\bm s = \chi^\dag \bm{\sigma} \chi$, where 
$\chi=(a,b)^\intercal / \sqrt{|a|^2+|b|^2}$ with $a$ and $b$ being the expansion coefficients of $\Psi$ in the HH ground states $|\pm 3/2 \rangle$. 
Blue and red arrows in Figs.~\ref{numerical_fig}(a2), \ref{numerical_fig}(b2), and \ref{numerical_fig}(c2) show the pseudospin orientations as a function 
of the in-plane wave vector $\bm k_{\parallel}$ for (001)-, (012)-, and (011)-grown QWs, respectively.  In the (011)-grown QW [Fig.~\ref{numerical_fig}(c2)], the pseudospin lies in the QW plane. In line with the symmetry consideration, the overall dependence of spin splitting on $\bm k_{\parallel}$ is described by the anisotropic Rashba Hamiltonian given by the first two terms in Eq.~\eqref{so_symmetry}. 
The striking difference of SO interaction in the (012)-grown QW [Fig.~\ref{numerical_fig}(c2)], compared to (001)- and (011)-grown structures, 
is that the pseudospin $\bm s$ has both the in-plane and out-of-plane components. Physically, it means that the eigen states are formed mostly 
from either $|+ 3/2 \rangle$ or $|- 3/2 \rangle$ states. 
In accordance with the symmetry analysis, such a kind of SO splitting 
is allowed in cubic QWs grown along low symmetry axes, see Fig.~\ref{structure_sketch}. Moreover, the numerical calculations show that, in the (012)-grown QW,  
$\bm s$ points mostly along the QW normal for $\bm k \parallel x$ where the spin splitting is maximal. All the observed features are well described by the effective Hamiltonian~\eqref{so_symmetry} with $|\zeta| > |\alpha_1|, |\alpha_2|$. 
  
\begin{figure}[!ht] 
	\centering
	\includegraphics[width=0.95\linewidth]{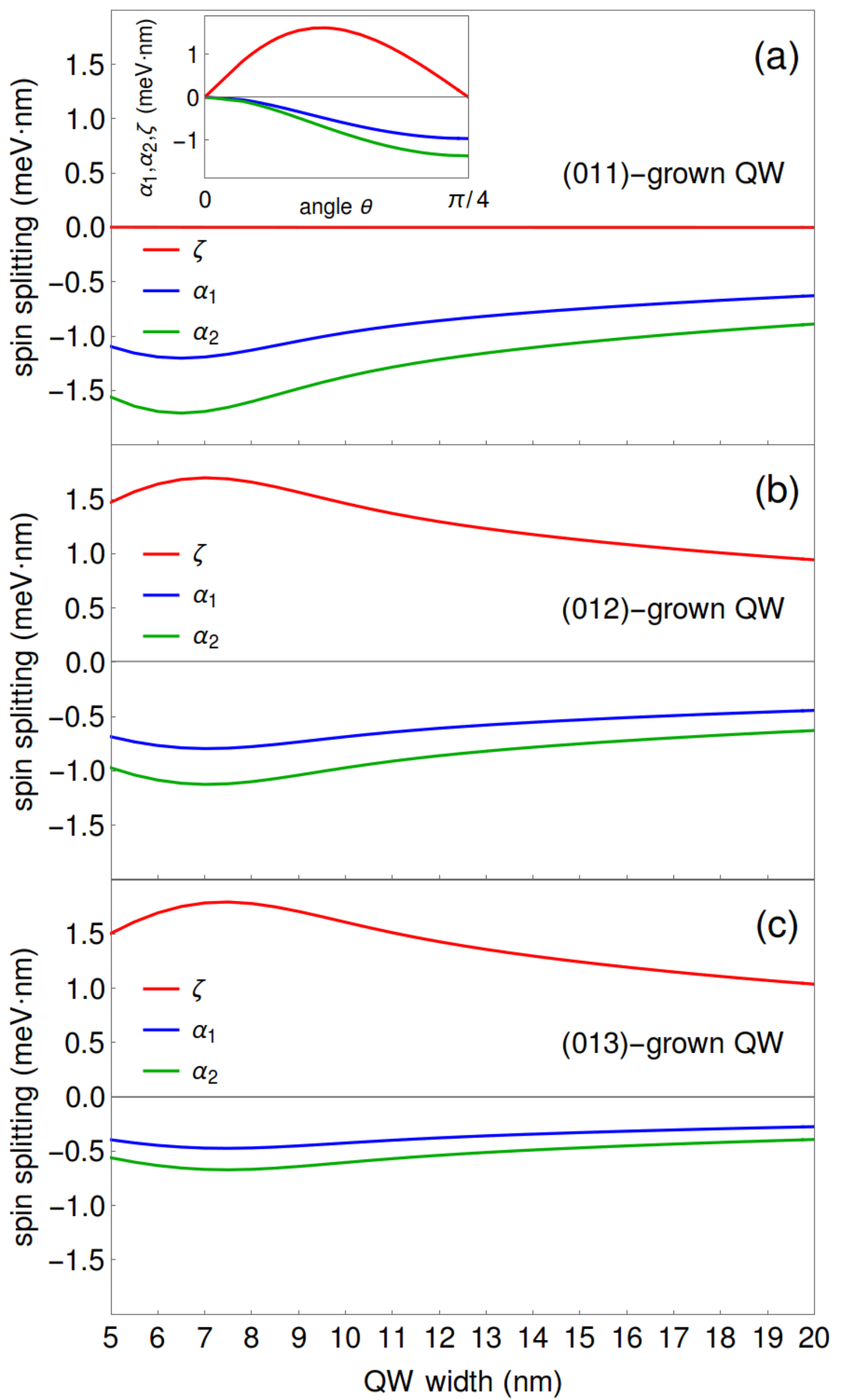} 
	\caption{
Dependences of the parameters of SO Hamiltonian~\eqref{so_symmetry} $\alpha_1$, $\alpha_2$, and $\zeta$ of heavy holes on QW width
for (011)-, (012)-, and (013)-grown GaAs/AlAs QWs. The dependences are calculated for the Luttinger parameters listed in Table~\ref{gamma_table} and built-in electric field $E_z=100$~kV/cm. Inset shows the dependence of $\alpha_1$, $\alpha_2$, and $\zeta$ on the growth direction $\theta$ for $10$-nm-wide QWs.
	} 
	\label{width_plot}
\end{figure}

We perform a series of numerical calculations to obtain the dependence of the SO coupling parameters $\alpha_1$, $\alpha_2$, and $\zeta$ on the 
the QW width and crystallographic orientation. The results are summarized in Fig.~\ref{width_plot}. The dependences of the SO parameters on the QW width turn out to be non-monotonic and have maxima at around $7$~nm. 
In narrower QWs, the influence of the electric field $E_z$, which is set to constant in the QW and zero in the barriers, on the HH states is reduced. 
As a result, the SO parameters decrease. In wide QWs, the SO parameters tend to constants 
since the well effectively becomes triangular with the slope $e E_z$.

Inset in Fig.~\ref{width_plot}(a) shows the dependence of the SO parameters on the angle $\theta$ which defines the QW crystallographic orientation. 
The angles $\theta=0$ and $\theta=\pi/4$ correspond to (001)-grown and (011)-grown QWs, respectively. In line with the symmetry consideration,
the parameters are zero at $\theta=0$ and, additionally, $\zeta = 0$ at $\theta=\pi/4$. The parameters $\alpha_1$ and $\alpha_2$ and proportional 
to $\theta^2$ at small $\theta$ and reach maxima at $\theta=\pi/4$, i.e., in (110)-grown QWs. The parameter $\zeta$ behaves differently: it is proportional to 
$\theta$ at small angles and reaches a maximum at $\theta \approx 0.36$ which is close to the (013) growth direction. The calculations show that, 
for GaAs/AlAs QWs, the term $\zeta \sigma_z k_x$ prevails in the SO Hamiltonian of heavy holes in a wide range of QW crystallographic orientations.
  
\section{Analytical theory}
\label{theory}  

Finally, we obtain analytical expressions for the SO coupling parameters. We assume that the warping of the valence-band spectrum, which is an essential ingredient for the $\bm k$-linear splitting of the HH subbands, is small, i.e., $|\gamma_2-\gamma_3| \ll \bar{\gamma}$, where $\bar{\gamma} = (\gamma_2 + \gamma_3)/2$. 

At $\bm k_\parallel = 0$  and in the absence of warping ($H_L^{({\rm a})} = 0$), 
solution of the Schr\"{o}dinger equation gives the series of the heavy-hole and light-hole states, 
$|HHn, \pm 3/2 \rangle$ and $|LHn, \pm 1/2 \rangle$, respectively,  with the energies $\varepsilon_{h/l,n}$ and the envelop functions 
$\varphi_{h/l,n}(z)$. 

At $\bm k_\parallel \neq 0$, the ground HH states $|HH1, + 3/2 \rangle$ and $|HH1, - 3/2 \rangle$ are coupled. 
The $\bm k$-linear coupling is described by the effective Hamiltonian~\eqref{so_symmetry}. The second-order perturbation theory
(first order in $\bm k_\parallel$ and first order in $\gamma_2 - \gamma_3$) gives the following equations for the SO coupling parameters 
(see Appendix A for details)
\begin{align}  
\label{theoryans}  
\alpha_{1,2} = \frac{3}{2} \frac{\hbar^2 Q}{m_0} \sin^2 2\theta \: , \;   \zeta= - \frac{3}{2} \frac{\hbar^2 Q}{m_0}  \sin 4\theta \:, 
\end{align}
where
\begin{multline}
\label{gequation}
Q = \frac{\hbar^2}{m_0} \sum_{n} \dfrac{ \int \varphi_{h,1}(z) \{\bar{\gamma},\partial_z\}_s \varphi_{l,n}(z) dz }{\varepsilon_{h,1} - \varepsilon_{l,n}}\\
\times \int  \varphi_{l,n}(z) \partial_z (\gamma_3-\gamma_2)\partial_z \varphi_{h,1}(z) dz \:.
\end{multline}
The parameter $Q$ is non-zero only if the QW is asymmetric and at least one of the Luttinger parameters has a jump at the interfaces 
(see Appendix B for details).

The dependences of $\alpha_1$, $\alpha_2$, and $\zeta$ on the growth direction $\theta$ calculated numerically in Fig.~\ref{width_plot} are
well described by Eqs.~\eqref{theoryans} despite the fact that the analytical theory is not strictly applicable to GaAs/AlAs QWs since
the difference between $\gamma_2$ and $\gamma_3$ is not small. As it follows from Eqs.~\eqref{theoryans}, the parameter $\zeta$ reaches a maximum at 
$\theta = \pi/8$ which is quite close to $0.36$ obtained in the numerical calculations.

\section{Summary} 
\label{summary}

To summarize, we have shown that the zero-field spin splitting of two-dimensional states in QWs grown along low-symmetry crystallographic axes 
goes beyond the Rashba and Dresselhaus contributions. In particular, there is spin-orbit coupling between the out-of-plane component of carrier's spin and the in-plane momentum whose strength is governed by QW structure inversion asymmetry. The developed microscopic theory shows 
that this type of coupling can be quite strong and dominate $\bm k$-linear spin splitting of heavy holes in $(0lh)$-oriented GaAs QWs. This finding 
provides an additional way to manipulate spins in low-dimensional structures.

More generally, our work suggests a novel type of tunable gyrotropy in low-symmetry QW structures: coupling between the out-of-plane component of an axial vector and the in-plane component of a polar vector controlled by structure inversion asymmetry. This coupling can give rise to a number of interesting phenomena, such as optical activity, magneto-electric effect, current-induced Faraday and Kerr rotation, circular photogalvanic effect, etc.

\section{Acknowledgments}
\label{acknow}
G.V.B. acknowledges the support from the Russian Science Foundation (project No.21-72-10035) and the ``BASIS'' foundation.

\section*{Appendix A}
         
The unperturbed Hamiltonian at $\bm{k}_\|=0$ is given by
\begin{equation}
\label{initial_hamiltonian}
H_0=H^{({\rm i})}_L+e V(z) \:.
\end{equation}
Here, we do not specify the shape of the electrostatic potential $V(z)$ and consider the general case.
The Hamiltonian~\eqref{initial_hamiltonian} is diagonal in the spin indices and gives the series of the two-fold degenerate 
heavy hole states $|HHn, \pm 3/2 \rangle$ and two-fold degenerate light hole states $|LHn, \pm 1/2 \rangle$ with the energies 
$\varepsilon_{h,n}$ and $\varepsilon_{l,n}$, respectively.
 
The perturbation Hamiltonian has the form ${\delta H = H-H_0}$, where $H$ is total Hamiltonian given by Eq.~\eqref{full_H}. Straightforward calculations show that the first-order corrections to the energy do not lead to $\bm k$-linear splitting. Thus, to calculate the splitting we use the second-order perturbation theory, which is first order in $\gamma_3-\gamma_2$ and first order in $\bm k_\parallel$. The linear in $\gamma_3-\gamma_2$ and
linear in $\bm k_\parallel$ contributions to $\delta H$ are given by
\begin{multline}
 R=\frac{2\hbar^2}{m_0} \left[\{J_y J_z\}_s \cos 2\theta + \frac{J_z^2 - J_y^2}{2} \sin 2\theta \right]\\\times \frac{k_z (\gamma_3-\gamma_2) k_z }{2} \sin 2\theta 
\end{multline}
and  
\begin{equation}
U =
\frac{2\hbar^2}{m_0} \biggl[\{J_x J_z\}_s \{\gamma_3, k_z \}_s\, k_x
+\{J_y J_z\}_s \{\gamma_2,k_z\}_s k_y\biggr] \:,
\end{equation}
 respectively. Using the standard perturbation theory we obtain the SO Hamiltonian of heavy holes in the basis of the states
 $| s\rangle = |HH1, + 3/2 \rangle$ and $| \bar{s} \rangle = |HH1, - 3/2 \rangle$
 \begin{equation}\label{H_secondorder}
H_{\text{SO}}=\sum_{\nu}
\dfrac{1}{\varepsilon_{h,1}-\varepsilon_{\nu}}{\small 
\begin{pmatrix}
R_{s \nu}U_{\nu s}+U_{s \nu} R_{\nu s} & U_{s \nu}R_{\nu \bar{s}}+R_{s \nu} U_{\nu \bar{s}}\\
U_{\bar{s} \nu}R_{\nu s}+R_{\bar{s} \nu} U_{\nu s} & R_{\bar{s} \nu}U_{\nu \bar{s}}+U_{\bar{s} \nu} R_{\nu \bar{s}}\\
\end{pmatrix} .
}
\end{equation}
Here, the index $\nu$ runs over all light-hole states $|LHn, \pm 1/2 \rangle$. The straightforward calculation of Eq.~\eqref{H_secondorder} 
yields the SO Hamiltonian~\eqref{so_symmetry} with the coupling parameters given by Eq.~\eqref{theoryans}.

\section{Appendix B}

If the Luttinger parameters $\gamma_i$ are independent on the $z$ coordinate, the integrals in Eq.~\eqref{gequation} can be simplified.
We note that the envelope functions $\varphi_{h,n}(z)$ and $\varphi_{l,n}(z)$ are the solutions of the equations
\begin{equation}  
\label{shrod}
-\dfrac{\hbar^2}{2 m_{h(l)}}\partial_z^2 \varphi_{h(l),n} = [\varepsilon_{h(l),n}-V(z)] \varphi_{h(l),n} \:, 
\end{equation}
where $m_h=m_0/(\gamma_1-2\gamma_2)$ and $m_l=m_0/(\gamma_1+2\gamma_2)$ are the heavy hole and light hole masses, respectively. 
Therefore, one has 
\begin{align}
\frac{\hbar^2}{2 m_h}\int \varphi_{l,n} \partial_z^2  \varphi_{h,1}  dz=\int \varphi_{l,n} [V(z)-\varepsilon_{h,1}] \varphi_{h,1} \:,\nonumber\\
\frac{\hbar^2}{2 m_l}\int \varphi_{l,n}\partial_z^2  \varphi_{h,1}  dz=\int \varphi_{l,n} [V(z)-\varepsilon_{l,n}] \varphi_{h,1} \:,
\end{align}
Subtracting the above equations from each other, one obtains
\begin{equation}
\label{integral_d_derivative}
\int \varphi_{l,n} \partial_z^2  \varphi_{h,1} dz = \frac{2}{\hbar^2} 
\dfrac{\varepsilon_{h,1}-\varepsilon_{l,n}}{1/m_l - 1/m_h}\int \varphi_{l,n}\varphi_{h,1} dz \:.
\end{equation}
With the help of Eq.~\eqref{integral_d_derivative} the parameter $Q$ given by the sum of integrals in Eq.~\eqref{gequation} 
can be presented in the form
\begin{multline}\label{Q_appendix}
Q=\dfrac{2 \bar{\gamma} (\gamma_3-\gamma_2) }{\dfrac{m_0}{m_l}-\dfrac{m_0}{m_h} }  
\sum_{n} \int \varphi_{h,1} \partial_z \varphi_{l,n} dz  \int \varphi_{l,n} \varphi_{h,1} dz' \:. 
\end{multline}
Finally, using the fact that the complete set of eigenfunctions satisfies the equation ${\sum_{n}\varphi_{l,n}(z)  \varphi_{l,n}(z')=\delta(z-z')}$,
the sum of integrals in Eq.~\eqref{Q_appendix} is further simplified to the integral $\int dz \: \varphi_{h,1} \partial_z  \varphi_{h,1}$
which is identically zero. Therefore, the SO coupling parameters $\alpha_{1,2}$ and $\zeta$ vanish if the Luttinger parameters 
in the heterostructure have no spatial dependence.

%\section*{References}  
\bibliography{scref6}

\end{document}